# Fractal carbon nanotube fibers with mesoporous crystalline structure

H. Yue[a]+, V. Reguero[a]+, E. Senokos[a,b], A. Monreal-Bernal[a], B. Mas[a], J.P. Fernández-Blázquez[a], R. Marcilla[b], J. J. Vilatela[a]*

[a] IMDEA Materials Institute, Getafe, 28906 Madrid, Spain

[b] IMDEA Energy Institute, Móstoles, 28935 Madrid, Spain

+These authors contributed equally to this work

**Abstract:** Macroscopic fibres of carbon nanotubes are hierarchical structures combining long building blocks preferentially oriented along the fibre axis and a large porosity arising from the imperfect packing of bundles. Synchrotron small-angle X-ray scattering SAXS measurements show that such structure is a surface fractal with fractal dimension ($D_s$) of 2.5 for MWCNT fibres and 2.8 for SWCNT fibres. N$_2$ adsorption measurements give similar values of 2.54 and 2.50, respectively. The fractal dimension and deviation from Porod's law are related to density fluctuations associated with the wide distribution of separations between CNTs. These fluctuations are also evident as diffuse wide-angle X-ray scattering (WAXS) from CNTs at distances above

---

* Corresponding author. Tel +34 915-493-422. E-mail: juanjose.vilatela@imdea.org (Juan J. Vilatela)

turbostratic separation. The structure of CNT fibres produced at different draw ratios is compared in terms of degree of orientation and characteristic lengths parallel and perpendicular to the fibre. Drawing not only increases alignment but also the fraction of graphitic planes forming coherent domains capable of taking part in stress transfer by shear; thus increasing both tensile modulus and strength. The invariant-normalized intensity of the (002) equatorial reflection thus takes the form of a degree of crystallinity closely related to tensile properties.

# 1. Introduction

The last decades have seen the development of a large library of fascinating nanostructured building blocks with wide range of chemical compositions and morphologies. Amongst the most studied of these are nanocarbons, 2D layers, transition metal di-chalcogenides, metal and metal oxide nanowires and cellulose nanofibers. In order to exploit their properties, these nanobuilding blocks are often assembled into macroscopic architectures and further processed to form composites, electrodes, membranes and other components. In the macroscopic ensemble the spatial distribution and interaction between building blocks play a critical part in determining the bulk properties of the system, often dominating over intra-particle features. The confluence of multiple length scale also implies that these macroscopic ensembles of nanobuilding blocks have a complex hierarchical structure that is both difficult to probe experimentally and to characterize precisely. Examples of different nanocarbon architectures include membranes produced by vacuum filtration (so-called bucky paper)[1], 3D porous scaffolds obtained by freeze-drying[2], and continuous macroscopic fibres of nanotubes[3,4,5] or graphene[6,7]. The case of macroscopic fibres of CNTs is particularly interesting for at least three reasons. The first stems from the possibility to adjust in a controlled way and independently the type of building block (in terms of the number of layers) and its orientation relative to the fibre axis[8]; with the added benefit that only one angle is required to describe the orientation between the 1D CNTs and the fibre axis. The second lies in the large evidence that bulk electrical and mechanical properties of these materials are strongly dependent on attributes of the CNT ensemble, such as CNT orientation and the degree of overlap, which determine on the one hand the effective inter-tube charge transfer resistance and on the other, the length over which stress is transferred in shear[9]. Piezoresistance[10], gas and liquid sensing[11] and capacitance[12] are also expected to depend on collective properties of the ensemble, such as specific surface area and pore size distribution. Few materials can be used both as a porous electrode as well as a high-performance reinforcing fibre, which itself points to an unusual structure. Thirdly, the production of these fibres in semi-industrial quantities[13] and fast pace of development application[14,3] gives testimony of their technological relevance. The hierarchical structure of CNT fibres has been studied mainly by electron microscopy analysis[15], including inspection of FIB-sections[16,17]. These techniques reveal the inner structure of the porous CNT fibres, but as a qualitative description constrained to a small sample volume. WAXS measurements have been used to determine the orientation of CNTs[18,19,15] and to characterize alignment heterogeneities[20]. SAXS studies have shown an intense component along the equator (fibre streak) attributed to elongated voids and bundles of CNTs. Micro-focus synchrotron studies showed a correspondence between CNT and bundle alignment, implying that there is no substantial misorientation of CNTs in the bundles, and used either the SAXS fibre streak or the (002) reflection to quantify fibre orientation[18]. Further analysis of SAXS was used to compare degrees of swelling and other relative structural changes in fibres after infiltration of liquids, based on the assumption that the scattering intensity strictly follows Porod's law for a two-phase system with sharp smooth boundaries[11,21].

Overall, a direct interpretation of SAXS data for systems far from an ideal multi-phase structure is challenging, and even more so when they are anisotropic[22]. That is the case for fibres of CNTs, and most likely for the other macroscopic ensembles of nanobuilding blocks mentioned above. The high porosity of CNT fibres makes them similar to porous carbons with a large rough internal area, such as activated carbon or glassy carbon, yet, consisting of high-aspect ratio pores preferentially orientated parallel to the fibres axis. Highly oriented regular needle-shaped pores are also present in carbon fibres (CF) and have been extensively studied by SAXS[23], however they are much smaller and present at significantly lower volume fractions.

This paper sets out as a multiscale-study of CNT fibres in an effort to gain further insight into their structure and bulk fibre properties. We treat the material as both a porous carbon and an oriented network of crystallites, and combine gas-adsorption and synchrotron WAXS-SAXS measurements. The results show that CNT fibres are in fact fractal surface structures, with features reminiscent of both CF and coal. The structural effects of applying higher draw ratios during fibre assembly in the gas-phase are analysed and related to tensile properties through a "degree of crystallinity" extracted from WAXS, which becomes a convenient predictor of tensile properties when comparing different CNT fibres.

## 2. Experimental

**2.1 Materials**

The CNT fibers were synthesized by the direct spinning method.[24] Briefly, the fibers are produced by continuous drawing of an aerogel of CNTs directly from the gas phase during CNT growth by chemical vapor deposition in a vertical reactor. Butanol, ferrocene and thiophene were used as carbon source, catalyst and promoter, respectively, in a concentration of 97.7:1.5:0.8 chosen to produce fiber made up of thin few-layer multiwall carbon nanotubes (MWCNT) with average diameter of 5 nm (S/C ratio = $3.3 \times 10^{-3}$ at. %). Fibres of predominantly singlewall carbon nanotubes (SWCNTs) were produced using a lower S/C ratio of $2.2 \times 10^{-4}$ at. %.

The draw ratio is defined as the elongation of the aerogel relative to the velocity of the carrier gas, defined as $DR = \frac{W}{V_{gas}}$, where $W$ is the rate at which the fibre is extracted from the reactor, i. e. the winding rate, and $V_{gas}$ the carrier gas velocity, typically ≈ 0.3 m/min. Winding rates are in the range 7 – 40 m/min.

## 2.2 Characterisation

Electron microscopy was carried out with a EVOR MA15 SEM using an accelerating voltage of 5~20 kV and a JEOL JEM 3000F TEM at 300kV. Gas adsorption measurements were carried out on a commercial surface area analyzer, Gemini VII 2390 (Micromeritics) and Quantachrome Quadrasorb SI porosimeter. Isotherm data (including adsorption and desorption) using nitrogen as the probe molecule were recorded at liquid nitrogen conditions, i.e., 77.3 K. Dried overnight CNT fibers (weighed 33-50 mg) were outgassed at 300 °C for 3 hours under vacuum prior to the measurement. Extensive precautions were taken to ensure that both the data and its mathematical treatment were accurate: samples of at least 33 mg were used for each measurement and measurements were repeated at least 2 times, Data we extracted from isotherms were accurate over the whole pressure range (SOM). To calculate specific surface area multi-point Brunauer–Emmett–Teller (BET) plot was employed within the adsorption branch of the isotherm in the relative pressure range $0.1 < P/P_0 < 0.3$, the surface fractal dimensions were calculated using

Frenkel-Halsey-Hill (FHH) method and pore size distribution was obtained using the Barrett-Joyner-Halenda (BJH) analysis applied to the nitrogen desorption isotherm data.

Simultaneous 2D WAXS/SAXS patterns of CNT fibers were collected at NCD beamline11, ALBA Synchrotron Light Facility, Spain. The radiation wavelength was 1.0 Å and the spot size at the focal plane of approximately 100 μm x 50 μm. Variations in incident beam intensity were negligible. The samples for WAXS/SAXS measurements consisted in tows of several filaments, typically 10-100, depending on the draw ratio applied at the point of spinning. The sample-to-detector distance and detector angle of tilt were determined using a reference material for calibration. Details of SAXS/WAXS data treatment are provided in supplementary material (SOM).

Tensile tests were carried out with a Favimat Tensile tester, using a gauge length of 20mm. The modulus was calculated from the tangent to the initial quasi-linear part of the stress-strain curve for strains < 0.5%.

## 3. Results and Discussion

A CNT fibre can be visualized as a network of bundles of nanotubes predominantly oriented parallel to the fibre axis but with a large porosity arising from the imperfect packing of the CNT bundles, as shown in the example in Fig. 1 for a sample of few-layer MWCNTs. In this work, the samples studied correspond to fibres produced by the direct spinning method, whereby an aerogel of CNTs is directly drawn out of the chemical vapour deposition chamber during growth of CNTs, but we expect the concepts developed to apply to various other macroscopic ensembles of nanobuilding blocks. The hierarchical structure of these fibres resembles that of staple yarns, made up subfilaments which in turn are made up of even smaller subfilmanets. Their tensile properties[25], knot resistance[26] and liquid uptake[27] are also reminiscent of yarn-like structures. At smaller scales, high magnification electron micrographs show that the pore structure is a reflection of a wide distribution of inter-bundle separations, as shown in Fig.1 c-d. Closed-packed bundles are in close proximity (<1 nm) and overlap along a fraction of their length, but then branch out and give rise to pores. The pores are thus delimited by the bundles and the pore sizes reflect interbundle separations. This pore structure is accessible by XRD measurements, which can probe from interatomic spacings (wide-angle) to mesopores (small-angle) while also taking into account the anisotropy of the sample, for example by using a 2D detector. The porous structure shown in Fig. 1 is also suitable for gas-adsorption studies, which cover the range 0.35 nm to tens of nm.

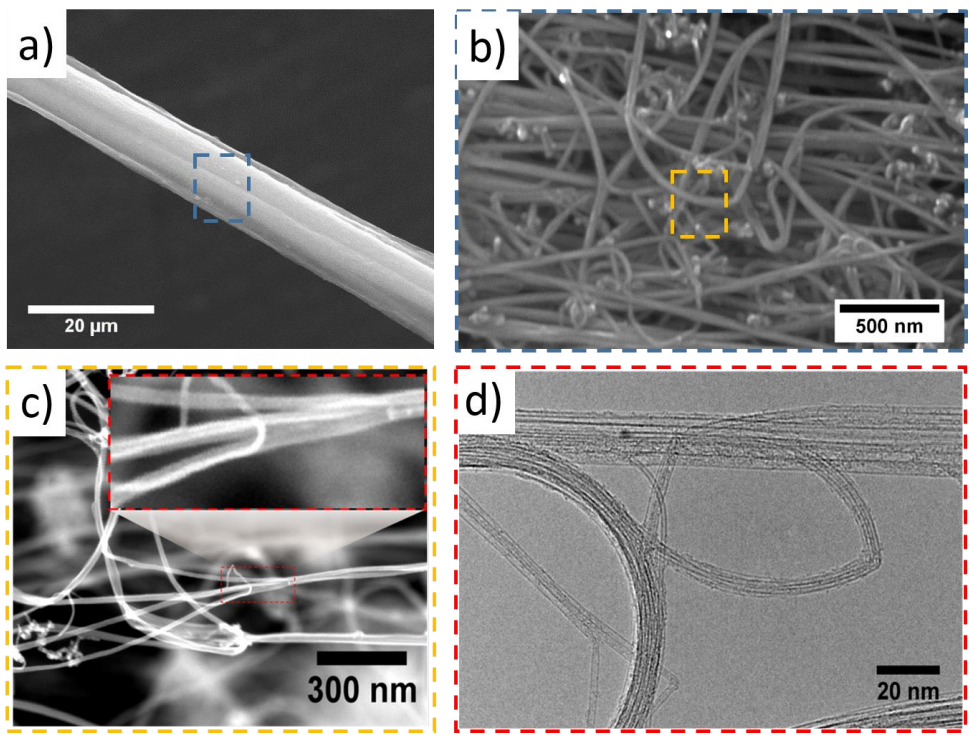

**Fig.1–Electron micrographs showing the hierarchical structure of CNT fibres. (a) Individual CNT fibre filament. (b) At low magnification the sample resembles a porous network (c). Higher resolution SEM micrographs show the mesoporous structure arising from imperfect**

**bundle packing (d) Pores are delimited by CNTs that branch out of bundles.3.1 CNT fibres as an isotropic porous material**

SAXS data for a typical CNT fibre sample is presented in Fig. 2a, consisting of a plot of scattering intensity against scattering vector *q*, obtained from full 360º azimuthal integration. The graph shows predominantly a smooth drop in intensity. In the Porod region towards higher q values ($q$=0.2~0.9 nm$^{-1}$, *d*-spacing about 7~30 nm), the power law $I(q) \propto q^{-K}$ gives a slope of $K = 3.4$. Materials with a uniform smooth surface are characterized by Porod slopes of 4, whereas values of $K < 4$ are typical of porous media, such as particle aggregates, colloidal systems, activated carbons and porous ceramics[28], which are better described as fractal structures. For $3 < K < 4$, as is the case for CNT fibres, the material corresponds to a surface fractal, instead of a mass fractal (for which K is < 3). For a surface fractal, $I(q) \propto q^{-(6-D_s)}$[29] and thus the surface fractal dimension, $D_s$ of CNT fibres, is 2.59.

Before providing an interpretation of the surface fractal dimension in this system, we introduce $N_2$ gas adsorption measurements, which probe the dimensionality of the CNT fibre array by a completely different mechanism, namely by adsorption of gas molecules on the material's surface. This method is based on adsorption isotherms using the modified Frenkel-Halsey-Hill (FHH) theory, according to

$$\ln(V/V_m) = \text{constant} + (D_{FHH} - 3)[\ln(\ln(P_0/P))]$$

where *V* is the volume of adsorbed gas at the relative pressure of $P/P_0$, $V_m$ the volume of monolayer coverage. Fig. 2b shows representative plots of $\ln(V/V_m)$ versus $\ln(\ln(P_0/P))$ of a CNT fibre. It is fairly linear over the entire adsorption range, although with two distinct regimes. The first one (regime I) is in the range -0.21≤ $\ln(\ln(P_0/P))$ ≤0.3 and corresponds to the Langmuir monolayer coverage of the CNTs by adsorbent gas molecules. The second (regime II) below $\ln(\ln(P_0/P))$ ≈- 0.21 corresponds to multilayer adsorption and capillary condensation. Only the monolayer coverage section of the adsorption isotherm provides accurate values for the dimensionality of the surface, as recently shown by Tang et al,[30] .Taking the slope in this regime (5.55), the resulting surface fractal dimension comes out as 2.54, which is nearly identical to the SAXS value.

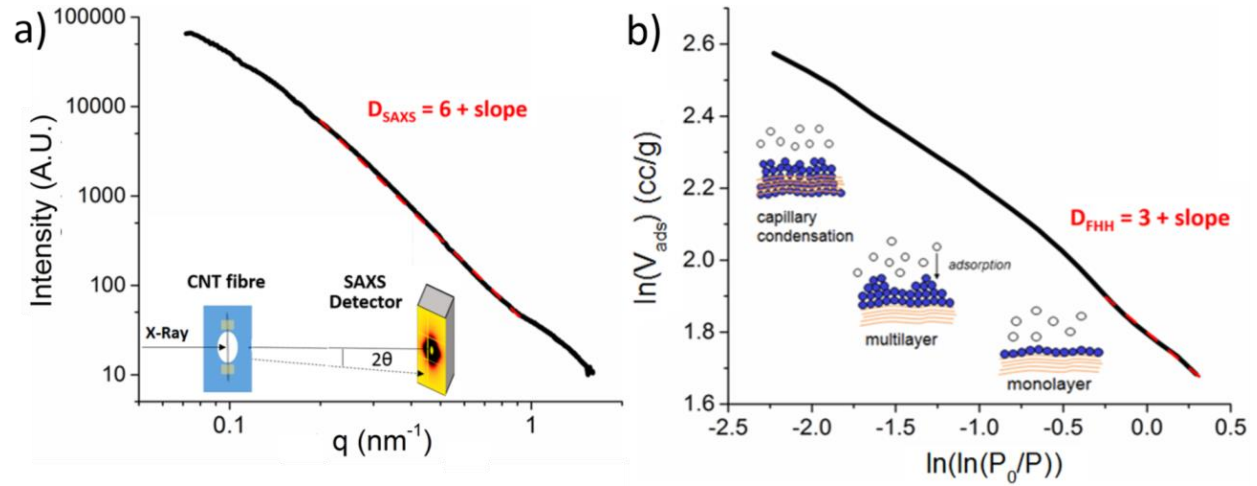

**Fig.2– Fractal dimension analysis of MWCNT fibres. (a) Plot of SAXS intensity against scattering vector. (b) Plot of nitrogen adsorption isotherm showing the linear fit in the regime where Langmuir monolayer adsorption is dominant before the capillary condensation occurs.**

The value of $D_s$ of 2.5-2.6 obtained is comparable to those reported for CNT membranes studied by SEM (2.43-2.86)[31], gas-adsorption alone (≈2.5)[32,33] and combined with 1D SAXS (2.6)[34] and dynamic light scattering (2.24-2.58)[35]. It is also in the range of values for activated carbon (2.8-3)[36] studied by the same techniques, nuclear graphite measured by small angle neutron scattering (2.5) and associated neutron scattering techniques extending up to the mm range (2.5)[37]. The convergence towards a value close to 2.5 is surprising considering differences in composition and expected pore size distribution of the different porous carbons. Nevertheless, there is considerable spread in values for different CNTs samples (2.43-2.86). This spread reflects morphological differences in the 3D aggregation or packing of nanotubes, which can arise from differences in the type of CNT and in their orientation. The diameter and number of layers of the CNTs determine their flexural rigidity and thus influence their ability to bend and reshape to maximise contact with adjacent tubes. The results in Fig. 2 correspond to fibres made up of MWCNTs with few (3.2+/-1.7 layers). Fibres of predominantly SWCNTs, produced by reducing the concentration of S promoter in the CVD reaction[38], give values of $D_s$ from SAXS 2.8 and $D_s$ from gas adsorption 2.5 (SOM).

**3.2 CNT fibres as an anisotropic porous material**

While the discussion above has treated the fibres as a porous carbon and found good agreement with similar materials in that group, the fibres are in fact highly anisotropic. Figs 3a and 3b present examples of 2D WAXS and SAXS patterns, showing a high degree of orientation of elements parallel to the fibre axis (vertical). The strong equatorial component at $q \approx 18.5$ nm$^{-1}$ corresponds to the (002) reflection from CNTs and includes contributions from internal layers of CNTs and inter-tube reflections (discussed further below). Its tail towards lower $q$ values reflects the distribution of separations between elements, which can also be interpreted as the openings of the elongated pores. Radial profiles for MCWNT and SWCNT fibres are included in Fig. 3c. The component at $q= 18.5$ nm$^{-1}$ corresponds to scattering from internal layers of CNTs as well as from CNT at turbostratic separations and thus represents graphitic layers in coherent domains. The lower $q$ component is due to the presence of CNTs at separations greater than the turbostratic distance. This scattering contribution is diffuse, somewhat similar to that of an amorphous polymer, and corresponds to CNTs that are too far apart to contribute significantly to stress transfer by shear. The oriented pore structure at larger spacings (pore sizes) is also evident in the 2D SAXS pattern. Taken as FWHM of the azimuthal profile after Lorentzian fitting, the degree of orientation remains fairly constant in the $q$ range 0.1 to 18 nm$^{-1}$ (Fig. 3d). We note that these values of FWHM include misalignment of filaments in the sample and thus underestimate the intrinsic orientation in the material (S2). Extracting values of fractal dimension, this time from equatorial or meridional integration instead of the full pattern, these values come out as 2.45 and 2.7 for MWCNT and 2.6 and 2.9 for SWCNT, respectively. They indicate that the fibres are rougher in the transverse direction, but show a weak dependence on orientation, at least in the range measured. A summary of values of fractal dimension values for different samples is provided in SOM, including data for fibres of SWCNTs.

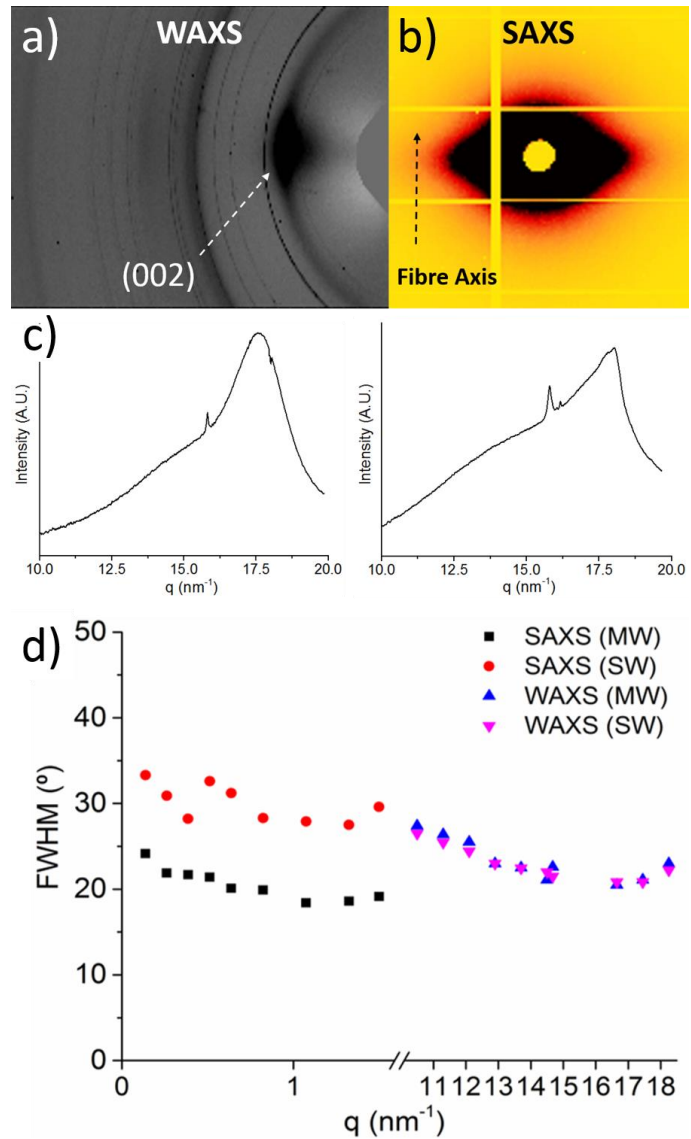

**Fig.3– Synchrotron data from CNT fibres oriented vertically. (a) 2D WAXS pattern showing catalyst powder rings and the oriented (002) reflection on the equator. (b) SAXS pattern with predominant equatorial orientation. (c) The radial profile along the equator confirms that the inter-tube scattering extends from the (002) at turbostratic separation up to $q \approx 10$ nm$^{-1}$. (d) The orientation of CNT fibres is fairly constant across different length-scales analysed by WAXS/SAXS.**

The fractal nature of these samples is closely linked with their network structure of graphitic planes in a rough disordered arrangement. Even graphite, with a much lower porosity, has a fractal surface across 6 orders of magnitude[39]. This is perhaps more intuitive in 2D maps obtained from neutron transmission in graphite, which produced irregular surfaces whose perimeter-area relation also obeys a power law with non-integer exponent and leading to a similar fractal dimension around 2.5 [39]. In its simplest interpretation, the fractal dimension of CNT fibres implies that they are not only porous, but that their effective surface is dependent on the size of the element used to measure it, i.e. of the $q$ range. Because these pores are open, they can be independently probed by gas adsorption measurements and thus provide the same fractal dimension. Their fractal dimension would seem to be linked to density fluctuations arising from imperfect packing of elements across the Å to ~100 nm length-scale. But unlike a standard porous carbon, in CNT fibres such density fluctuations are anisotropic, with the pores elongated in the fibre direction. Needle-shaped pores and density fluctuations have been extensively studied in CF[40] and glassy carbon[41] and thus provide an interesting comparison. The distribution of interlayer spacings in graphitic carbons produces density fluctuations that lead to non-integer Porod slopes (1.85 to 2.58 in Kratky slit optics[42]) that could be interpreted as fractal structure in the sense that the effective surface of scattering is dependent on $q$. For CNT fibres the dominant porosity is inter-particle, arising from imperfect packing of CNTs. The internal "porosity" of the hollow CNTs (intra) is expected to be more regular and, because of the small diameter < 3nm of the CNTs in this study, therefor only observed at relatively large q values. A small peak at q ~ 1 nm$^{-1}$ in equatorial profiles has been previously attributed to the CNT hollow cylinder form factor,[43] and indeed it is observed in aligned CNT fibres (SOM).

A fractal dimension of around 2.5 is in fact common in various disordered systems with interconnected and ramified porosity, more generally defined as fracturing ranked surfaces[44]. Such structures arise in graphitic materials most likely because of the high anisotropy of graphitic domains, where the high in-plan rigidity and weak interaction between layers inherently favour the formation of disordered porous systems by imperfect packing. This effect is magnified in CNT systems because of their enormous aspect ratio, in the range of $10^5 - 10^6$ for the CNTs in this work.

**3.3 Structure-properties**

It is of interest to relate specific structural features of the CNT fibre to bulk fibre properties, and particularly to be able to control the fibre structure during its production in order to tailor it and thus obtain desired properties. With this objective, we have analysed a series of CNT fibres with different draw ratios. Fibres with low degree of alignment (DR<70) have tensile properties in the range of a polyester though nevertheless superior to most metals on a mass basis (i.e. normalized by specific gravity), whereas highly oriented fibres (DR≈130) of the same molecular composition have properties in the high-performance range approaching aramid fibre (Kevlar). Increasing the DR produces, for example, a variation of tensile modulus from 18 GPa/SG to 62 GPs/SG and an increase in specific strength from 0.3 GPa/SG to 1.1 GPa/SG. The structural features that give rise to these differences in tensile properties are not captured by changes in the fractal dimension (Table S2), but in finer details of the WAXS-SAXS and the pore size distribution obtained by gas-adsorption. This is not surprising considering that tensile and electrical properties depend on load and charge transfer processes across CNTs over interatomic potential distances < 1nm, and thus not sensitive to porosity over multiple length-scales. Table 1 presents tensile properties and different structural parameters for few-layer MWCNT fibre samples produced with different DR. $L_p$, the Porod chord length, is the average lateral size of the pore; $L_3$, the average length of pore; $I_{(002)}$ is intensity of the (002) normalized by the invariant. The width of mesopore size distribution is extracted from N$_2$ adsorption measurements. Details are included in SOM4.

**Table 1. Structure-properties of CNT fibres produced at different draw ratios.**

| Draw ratio | SAXS | | | WAXS | | FWHM of BJH pore size distribution (nm) | Strength (GPa/SG) | Modulus (GPa/SG) |
|---|---|---|---|---|---|---|---|---|
| | FWHM (°) | $L_p$ (nm) | $L_3$ (nm) | FWHM (°) | $I_{(002)}$ | | | |
| 4.7 | 52 | 20.3 | 159 | 35 | - | 33 | 0.2 | 3.4 |
| 31.5 | 31 | 17.6 | 164 | 21 | 28.2 | - | 0.3 | 18 |
| 63 | 25 | 17.3 | 210 | 21 | 37.5 | - | 0.65 | 32.3 |
| 95 | 25 | 13.8 | 227 | 19 | 40.8 | - | 0.82 | 38.5 |
| 126 | 26 | 13.1 | 221 | 18 | 48.6 | 44 | 1.1 | 62.4 |

As expected, drawing improves alignment of the CNTs. This has the additional effect of making the pores narrower (lower $L_p$) and more elongated (higher $L_3$), as well as narrowing the distribution of pore sizes (w). The schematic in Fig. 4 illustrates the structure of fibres produced at high DR, according to the SAXS and gad-adsorption data in Table 1. Waviness can be ignored on this length-scale because of the high persistence length of CNTs > 100nm.

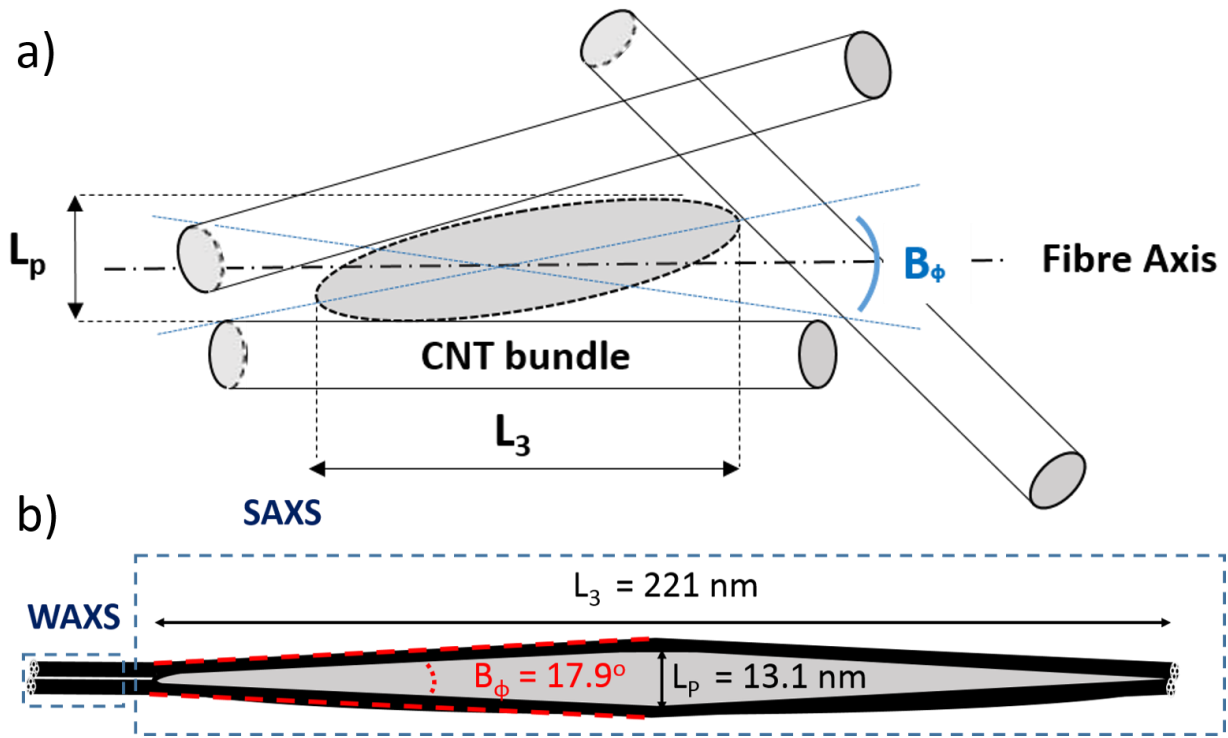

**Fig.4–Fibre structure according SAXS results. (a) Parameters determined from 2D SAXS and their assignation. (b) Schematic showing the resulting fibre structure for a highly drawn sample (DR = 126).**

More importantly, the WAXS results in Table 1 indicate that applying higher DR to the CNT fibre at the point of assembly in the gas-phase not only reorients the CNTs and bundles, but additionally leads to an increase in the fraction of CNTs at turbostratic separations. Fig. S4a presents a comparison of diffraction patterns, which clearly show that the more oriented materials has a stronger (002) equatorial reflection. The normalized intensity of this component $I_{(002)}$ increases with increasing DR (Fig. S4b), implying that a larger fraction of the material consists of graphitic layers at sufficiently close separation to transfer stress by shear. The cut off distance for mechanical interaction between graphitic planes is *a priori* unknown, but it is clear that the combined increase in tensile strength and modulus is largely due to this improvement in bundle packing. An increase in tensile modulus could in principle be ascribed entirely to improved orientation of crystalline domains, following well established mechanical models for high-performance fibres[45],[46]. But the increase in tensile strength must originate from the formation of new shear load-bearing interfaces upon drawing, which thus provide an increment in maximum load per unit mass that the fibre can withstand before failure. In this respect, CNT fibres would seem to behave like molecular solids, rather than monolithic crystalline materials, with the coexistence of non-coherent and coherent domains and the integral of the (002) reflection normalized by the invariant represents a measure of degree of effective crystallinity for stress transfer.

## Conclusions

CNT fibres combine structural features of porous carbon and of high-performance fibres. The imperfect packing of CNTs, produces a large volume of elongated mesopores delimited by the network of CNT bundles. Thus, pore openings are equivalent to where CNTs branch out of common bundles. This bifurcation of CNTs is also observed as a wide distribution of inter-CNT spacings extending well beyond turbostratic separation. The resulting hierarchical structure can be described as a surface fractal, with similar fractal dimensions obtained by SAXS and by $N_2$ adsorption around 2.5. The samples are rougher in the transverse direction, with fractal dimensions of 2.45 and 2.6 for fibres of MWCNT and SWCNT, respectively, compared with 2.7 and 2.9 for the longitudinal direction.

These concepts were applied to samples produced at different draw ratios in order to determine the structural features that arise at the point of CNT assembly in the gas phase that later determine, for example, tensile properties. SAXS shows that pores become more oriented, sharper and more elongated as the fibre is drawn. But the improvement in orientation for higher draw ratios has the additional effect of increasing the fraction of graphitic elements at sufficiently close separation to take part in stress transfer and as a result both modulus and tensile strength of the bulk fibre increase. This resembles the behaviour of high-performance polymer fibres. Heat-tensioning of rigid-rod polymer fibres, for example, has the dual effect of improving orientation and increasing lateral coherence size, thus doubling both tensile strength and modulus[47]. In the case of CNT fibres, determining a coherent domain size from WAXS is challenging because of the contribution from internal layers of CNTs and the wide distribution of inter-tube spacings. With respect to the latter, simulations show that the energy barrier for translational displacement in turbostratic graphite is very small[48] and it is therefore expected that negligible stress transfer takes place across intertube distances >~ 0.5 nm, but a precise cut-off distance is still to be determined.

## ASSOCIATED CONTENT

**Supporting Information**. Details of $N_2$ adsoprtion measurements, misorientation in multifilament samples, fractal dimension values for different samples (including SWCNT fibres), sstructural parameters of CNT fibres extracted from SAXS and gas adsorption data, and "Degree of crystallinity": fraction of graphitic planes taking part in stress transfer obtained by normalising the equatorial intensity of the (002) by the invariant. This material is available free of charge via the Internet at http://pubs.acs.org.

## AUTHOR INFORMATION

**Corresponding Author**

* E-mail:juanjose.vilatela@imdea.org18

**Author Contributions**

The manuscript was written through contributions of all authors. All authors have given approval to the final version of the manuscript.

# ACKNOWLEDGMENTS

Financial support is acknowledged from the European Union Seventh Framework Program under grant agreements 678565 (ERC-STEM), and FP7-People-Marie Curie Action-CIG (2012-322129 MUFIN), and from MINECO (MT2012-37552-C03-02, MAT2015-62584-ERC, RyC-2014-15115, Spain) and CAM MAD2D project (S2013/MIT-3007). Synchrotron XRD experiments were performed at NCD beamline at ALBA Synchrotron Light Facility with the collaboration of ALBA staff. The authors are grateful to Dr. P. Pizarro for assistance with interpretation of gas adsorption measurements and to Mr. J.C. Fernández for mechanical testing of fibres. JJV is grateful to Dr. J. Elliott for discussions about interlayer stress transfer in graphitic systems.

# REFERENCES

1. Rinzler, A. G. *et al.* Large-scale purification of single-wall carbon nanotubes: process, product, and characterization. *Appl. Phys. Mater. Sci. Process.* **67,** 29–37 (1998).
2. González, M., Baselga, J. & Pozuelo, J. High porosity scaffold composites of graphene and carbon nanotubes as microwave absorbing materials. *J Mater Chem C* **4,** 8575–8582 (2016).
3. Liu, Y. & Kumar, S. Polymer/Carbon Nanotube Nano Composite Fibers–A Review. *ACS Appl. Mater. Interfaces* **6,** 6069–6087 (2014).

4. Lu, W., Zu, M., Byun, J.-H., Kim, B.-S. & Chou, T.-W. State of the Art of Carbon Nanotube Fibers: Opportunities and Challenges. *Adv. Mater.* **24,** 1805–1833 (2012).

5. Vilatela, J. J. & Eder, D. Nanocarbon Composites and Hybrids in Sustainability: A Review. *ChemSusChem* **5,** 456–478 (2012).

6. Xu, Z. & Gao, C. Graphene fiber: a new trend in carbon fibers. *Mater. Today* **18,** 480–492 (2015).

7. Li, Z., Liu, Z., Sun, H. & Gao, C. Superstructured Assembly of Nanocarbons: Fullerenes, Nanotubes, and Graphene. *Chem. Rev.* **115,** 7046–7117 (2015).

8. Alemán, B., Reguero, V., Mas, B. & Vilatela, J. J. Strong Carbon Nanotube Fibers by Drawing Inspiration from Polymer Fiber Spinning. *ACS Nano* **9,** 7392–7398 (2015).

9. Vilatela, J. J., Elliott, J. A. & Windle, A. H. A Model for the Strength of Yarn-like Carbon Nanotube Fibers. *ACS Nano* **5,** 1921–1927 (2011).

10. Lekawa-Raus, A., Koziol, K. K. K. & Windle, A. H. Piezoresistive Effect in Carbon Nanotube Fibers. *ACS Nano* **8,** 11214–11224 (2014).

11. Qiu, J. *et al.* Liquid Infiltration into Carbon Nanotube Fibers: Effect on Structure and Electrical Properties. *ACS Nano* **7,** 8412–8422 (2013).

12. Senokos, E., Reguero, V., Palma, J., Vilatela, J. J. & Marcilla, R. Macroscopic fibres of CNTs as electrodes for multifunctional electric double layer capacitors: from quantum capacitance to device performance. *Nanoscale* **8,** 3620–3628 (2016).

13. Schauer, M. W. & White, M. A. Tailoring Industrial Scale CNT Production to Specialty Markets. *MRS Proc.* **1752,** (2015).

14. Vilatela, J. J. & Marcilla, R. Tough Electrodes: Carbon Nanotube Fibers as the Ultimate Current Collectors/Active Material for Energy Management Devices. *Chem. Mater.* **27,** 6901–6917 (2015).

15. Beese, A. M. *et al.* Key Factors Limiting Carbon Nanotube Yarn Strength: Exploring Processing-Structure-Property Relationships. *ACS Nano* **8,** 11454–11466 (2014).

16. Wu, A. S. *et al.* Carbon nanotube fibers as torsion sensors. *Appl. Phys. Lett.* **100,** 201908 (2012).

17. Mora, R. J., Vilatela, J. J. & Windle, A. H. Properties of composites of carbon nanotube fibres. *Compos. Sci. Technol.* **69,** 1558–1563 (2009).

18. Davies, R. J., Riekel, C., Koziol, K. K., Vilatela, J. J. & Windle, A. H. Structural studies on carbon nanotube fibres by synchrotron radiation microdiffraction and microfluorescence. *J. Appl. Crystallogr.* **42,** 1122–1128 (2009).

19. Behabtu, N. *et al.* Strong, Light, Multifunctional Fibers of Carbon Nanotubes with Ultrahigh Conductivity. *Science* **339,** 182–186 (2013).

20. Pichot, V. *et al.* X-ray microdiffraction study of single-walled carbon nanotube alignment across a fibre. *Europhys. Lett. EPL* **79,** 46002 (2007).

21. Terrones, J., Elliott, J. A., Vilatela, J. J. & Windle, A. H. Electric Field-Modulated Non-ohmic Behavior of Carbon Nanotube Fibers in Polar Liquids. *ACS Nano* **8,** 8497–8504 (2014).

22. Stribeck, N. Extraction of domain structure information from small-angle scattering patterns of bulk materials. *J. Appl. Crystallogr.* **34,** 496–503 (2001).

23. Perret, R. & Ruland, W. The microstructure of PAN-base carbon fibres. *J. Appl. Crystallogr.* **3,** 525–532 (1970).

24. Li, Y.-L. Direct Spinning of Carbon Nanotube Fibers from Chemical Vapor Deposition Synthesis. *Science* **304,** 276–278 (2004).

25. Vilatela, J. J., Elliott, J. A. & Windle, A. H. A Model for the Strength of Yarn-like Carbon Nanotube Fibers. *ACS Nano* **5,** 1921–1927 (2011).

26. Vilatela, J. J. & Windle, A. H. Yarn-Like Carbon Nanotube Fibers. *Adv. Mater.* **22,** 4959–4963 (2010).

27. Neimark, A. V. *et al.* Hierarchical Pore Structure and Wetting Properties of Single-Wall Carbon Nanotube Fibers. *Nano Lett.* **3,** 419–423 (2003).

28. Brinker, C. J. & Scherer, G. W. *Sol-gel science: the physics and chemistry of sol-gel processing*. (Academic Press, 1990).

29. Roe, R. J. *Methods of X-ray and neutron scattering in polymer science*. (Oxford University Press, 2000).

30. Tang, P., Chew, N. Y. K., Chan, H.-K. & Raper, J. A. Limitation of Determination of Surface Fractal Dimension Using $N_2$ Adsorption Isotherms and Modified Frenkel−Halsey−Hill Theory. *Langmuir* **19,** 2632–2638 (2003).

31. De Nicola, F. *et al.* Multi-Fractal Hierarchy of Single-Walled Carbon Nanotube Hydrophobic Coatings. *Sci. Rep.* **5,** 8583 (2015).

32. Kanyó, T. *et al.* Quantitative Characterization of Hydrophilic−Hydrophobic Properties of MWNTs Surfaces. *Langmuir* **20,** 1656–1661 (2004).

33. Smajda, R., Kukovecz, Á., Kónya, Z. & Kiricsi, I. Structure and gas permeability of multi-wall carbon nanotube buckypapers. *Carbon* **45,** 1176–1184 (2007).

34. Sun, C.-H., Li, F., Ying, Z., Liu, C. & Cheng, H.-M. Surface fractal dimension of single-walled carbon nanotubes. *Phys. Rev. B* **69,** (2004).

35. Khan, I. A. *et al.* Fractal structures of single-walled carbon nanotubes in biologically relevant conditions: Role of chirality vs. media conditions. *Chemosphere* **93,** 1997–2003 (2013).

36. Pfeifer, P. *et al.* Nearly Space-Filling Fractal Networks of Carbon Nanopores. *Phys. Rev. Lett.* **88,** (2002).

37. Zhou, Z. *et al.* From nanopores to macropores: Fractal morphology of graphite. *Carbon* **96,** 541–547 (2016).

38. Reguero, V., Alemán, B., Mas, B. & Vilatela, J. J. Controlling Carbon Nanotube Type in Macroscopic Fibers Synthesized by the Direct Spinning Process. *Chem. Mater.* **26,** 3550–3557 (2014).

39. Zhou, Z. *et al.* From nanopores to macropores: Fractal morphology of graphite. *Carbon* **96,** 541–547 (2016).

40. Perret, R. & Ruland, W. Single and multiple X-ray small-angle scattering of carbon fibres. *J. Appl. Crystallogr.* **2,** 209–218 (1969).

41. Perret, R. & Ruland, W. X-ray small-angle scattering of glassy carbon. *J. Appl. Crystallogr.* **5,** 183–187 (1972).

42. Perret, R. & Ruland, W. X-ray small-angle scattering of non-graphitizable carbons. *J. Appl. Crystallogr.* **1,** 308–313 (1968).

43. Meshot, E. R. *et al.* Engineering Vertically Aligned Carbon Nanotube Growth by Decoupled Thermal Treatment of Precursor and Catalyst. *ACS Nano* **3,** 2477–2486 (2009).

44. Schrenk, K. J., Araújo, N. A. M., Andrade Jr, J. S. & Herrmann, H. J. Fracturing ranked surfaces. *Sci. Rep.* **2,** (2012).

45. Northolt, M. G., Veldhuizen, L. H. & Jansen, H. Tensile deformation of carbon fibers and the relationship with the modulus for shear between the basal planes. *Carbon* **29,** 1267–1279 (1991).

46. Young, R. J. & Eichhorn, S. J. Deformation mechanisms in polymer fibres and nanocomposites. *Polymer* **48,** 2–18 (2007).

47. Allen, S. R., Farris, R. J. & Thomas, E. L. High modulus/high strength poly-(p-phenylene benzobisthiazole) fibres: Part 2 Structure-property investigations. *J. Mater. Sci.* **20,** 4583–4592 (1985).

48. Shibuta, Y. & Elliott, J. A. Interaction between two graphene sheets with a turbostratic orientational relationship. *Chem. Phys. Lett.* **512,** 146–150 (2011).